\documentclass[12pt]{JHEP3}
\usepackage{epsfig}
\usepackage{amsmath}
\title{
Holographic Nuclei
}
\author{
Koji Hashimoto\\
Theoretical Physics Lab.,  
Nishina Center, 
RIKEN, Saitama 351-0198, Japan\\
E-mail: \email{koji@riken.jp}\\
}

\abstract
{We provide a dual gravity description of heavy atomic nuclei,
via AdS/CFT correspondence. In holographic QCD such as Sakai-Sugimoto
model, baryons are D-branes wrapping a sphere in 10 dimensional curved 
spacetime, so any nucleus is a collection of $A$ such D-branes where 
$A$ is mass number of the nucleus. 
Quantum theory on the nucleus is ADHM-like $U(A)$ Yang-Mills-Higgs 
theory on
the sphere. Taking a large $A$ limit (corresponding to heavy nuclei)
leads to a dual gravity describing collective excitataions of
constituent nucleons of the heavy nucleus. This dual gravity computes
spectra of the heavy nucleus, and gives discrete states 
whose gap roughly agrees with experimental nuclear data. 
}

\preprint{
{\normalsize RIKEN-TH-141}
}

\begin{document}

\section{From superstring to nuclear theory}
\label{section1}

Application of a superstring technique, the 
gauge/gravity (AdS/CFT) correspondence \cite{Maldacena:1997re,
Gubser:1998bc}, 
to low energy QCD has provided 
quite remarkable progress on hadron physics. This subject called
holographic QCD has grown up to be a major research arena of string
theory. Dual gravity description has revealed various aspects of low
energy QCD which were unreachable by conventional analytic methods
because of notorious strong coupling. 
They include spectra of glueballs, mesons and
baryons, and interactions among them, and even phase transitions at
finite temperature/density. However, almost all of the results are 
on hadron physics,\footnote{Inclusion of baryon chemical potential or
finite baryon density is a hot topic
in holographic QCD, see \cite{ref} for a partial list of the
references.}  
not really nuclear physics, dare to mention. In this paper, we take a
first step toward 
nuclear physics: we provide a dual gravity description of
{\it heavy nuclei}. 

The essence of the gauge/gravity correspondence is the large $N$ limit,
where $N$ refers to the rank of the gauge group $U(N)$ of the gauge
theory living on $N$ coincident D-branes. This limit, together with
large $\lambda$ ('tHooft coupling) limit, allows a dual, holographic,
equivalent gravity theory on a near horizon geometry of black brane 
solutions created by the $N$ D-branes. 

Baryons in QCD-like gauge
theories are described, in the dual gravity description,  
by in fact additionally put D-branes wrapping a higher dimensional sphere
in 10 dimensional curved spacetime \cite{Gross-Ooguri}. 
Any nucleus is a collection of baryons (nucleons), 
so in holographic QCD the nucleus 
is a collection of D-branes. Therefore,  
heavy nuclei with large mass number $A$ 
can have a dual gravity description provided by a near horizon geometry
of those ``baryonic'' $A$ D-branes, in large $A$ limit.
In this paper we pursue and realize this idea. We 
describe a possible dual geometry, check
its validity as a limit of string theory, and provide a comparison with
nuclear data. 

In the next section, we describe 
a gauge theory on the $A$ D-branes representing the
nucleus, with a brief review of baryons in AdS/CFT. 
Then in Sec.~3, we take a large $A$ limit and give a gravity dual, whose
fluctuation analysis is discussed in Sec.~4 to provide spectra of
heavy nuclei. The last section is for a
conclusion and discussions on possible relations to shell model and 
liquid drop model of nuclei, and black holes.

\section{Gauge theory on nucleus}

Among various holographic models of QCD, 
Sakai-Sugimoto (SS) model \cite{SaSu1} 
is a successful one given by top-down approach from string theory. 
We will work in this SS model. Baryons were 
introduced in the model in \cite{HSSY, Hong-Rho-Yee-Yi},
while in the context of AdS/CFT correspondence 
description of baryons in terms of D-branes 
has a much longer history \cite{Gross-Ooguri}.  

For ${\cal N}=4$ $U(N)$ 
supersymmetric Yang-Mills theory, the gravity dual is 
type IIB supergravity theory on $AdS_5\times S^5$. Baryon is described
by a D5-brane wrapping the $S^5$. All the spatial worldvolume 
dimensions of the D5-brane are in this ``internal'' $S^5$, thus it 
is point-like in the $AdS_5$. The reason why this D5-brane is
a baryon is that this supergravity has a background 5-form field
strength penetrating the $S^5$, with $N$ units of flux. 
This induces electric charge $N$
on the spherical D5-brane. Consistency of the charge conservation
requires fundamental strings emanating from the D5-brane surface
since electric charges are created at the end points of fundamental
strings on a D-brane. The strings ending on the D5-brane should be
elongated to the AdS boundary, and those are identified as 
(infinitely heavy) quarks. Thus the spherical D5-brane
consists of $N$ quarks, that is, a baryon. Precise supersymmetric
configuration of the D5-brane with the electric charge was given in 
\cite{Imamura:1998gk}.

One can collect $A$ D5-branes on top of each other. On the worldvolume
($\sim {\bf R}\times S^5$) of the $A$ D5-branes, 
as usual for any D-brane systems in string theory,
there appears a $U(A)$ super 
Yang-Mills theory. It includes scalar fields in
adjoint representation which measure relative position of the D5-branes.
Once the $S^5$ worldvolume is 
decomposed into spherical harmonics, the lowest mode depends effectively 
only on time, so the theory reduces to a $U(A)$ 
supersymmetric matrix quantum mechanics. 

The situation is qualitatively similar for the SS model.
The model provides various interesting features of baryons 
\cite{HSSY,Hong-Rho-Yee-Yi,PaYi,HSS}.
As for the gravity dual of
the gluonic part of QCD, the model adopts so-called Witten's geometry
\cite{Witten:D4}, 
\begin{eqnarray}
&& 
ds^2 = \left(\frac{U}{R}\right)^{3/2}
\left(
\eta^{\mu\nu}dx^\mu dx^\nu + f(U)d\tau^2
\right)
 + 
\left(\frac{R}{U}\right)^{3/2} 
\left(
\frac{dU^2}{f(U)} + U^2 d\Omega_4^2
\right),
\label{g}
\\
&& e^{\phi} = g_s \left(\frac{U}{R}\right)^{3/4},\quad
f(U)\equiv 1-\frac{U_{\rm KK}^3}{U^3},
\end{eqnarray}
with a 4-form field strength flux penetrating the $S^4$ specified by the
volume element $d\Omega_4$ above.
This is a supersymmetry-breaking solution of type IIA supergravity in 10
dimensions. The supergravity parameters are related to QCD parameters as
\begin{eqnarray}
 R = \left(
\frac{g_{\rm YM}^2 N_c l_s^2}{2 M_{\rm KK}}
\right)^{1/3}, \quad U_{\rm KK} = \frac29 g_{\rm YM}^2 
N_c M_{\rm KK} l_s^2, \quad g_S = 
\frac{g_{\rm YM}^2}{2\pi M_{\rm KK} l_s}.
\label{sugrap}
\end{eqnarray}
Flavors are introduced as $N_f$ D8-branes in this geometry. The
D8-branes extend all spatial directions except for $\tau$ parameterizing
$S^1$ whose radius is $1/M_{\rm KK}$. All the hadronic physical
observables computed in this model 
are given as functions of $N_c, g_{\rm YM}$ and $M_{\rm KK}$. Because
the geometry (\ref{g}) is truncated smoothly at $U=U_{\rm KK}$, 
one can introduce a new coordinate $(z,y)$ instead of $(\tau,U)$, 
as $z + iy = \sqrt{(U^3/U_{\rm KK})-U_{\rm KK}^2} \exp(iM_{\rm KK}\tau)$,
then the D8-brane location is expressed as $y=0$. 
The D8-brane gauge group $U(N_f)$
corresponds to unbroken global symmetry in QCD, the vector part of
the chiral symmetry.

The baryons in the SS model are given by D4-branes wrapping the $S^4$
\cite{SaSu1}, in the gravity dual side. See the table below, and
Fig.~\ref{fig}. It is particle-like and 
localized in the $(x^1,x^2,x^3)$ space.
\begin{center}
\begin{tabular}{c|c|c|c|c|c|c|c}
 &0 & 1& 2 & 3 & $y$ & $z$ & $S^4$
\\\hline
$N_f$ D8s & $\circ$ & $\circ$ & $\circ$ & $\circ$ & &$\circ$ &$\circ$
\\\hline
$A$ D4s & $\circ$ &  &  &  & & & $\circ$
\end{tabular}
\end{center}
The D4-branes sitting on the flavor D8-branes can be represented by 
Yang-Mills instantons (localized in $(x^1,x^2,x^3,z)$ directions) 
of the D8-brane gauge theory
\cite{brane-within-brane}, and in fact, due to a Chern-Simons coupling on
the D8-branes, the Yang-Mills instanton
sources an overall $U(1)$ gauge field on the D8-branes 
\cite{HSSY} and thus 
possesses a baryon number.

\FIGURE[t]{
\includegraphics[width=7cm]{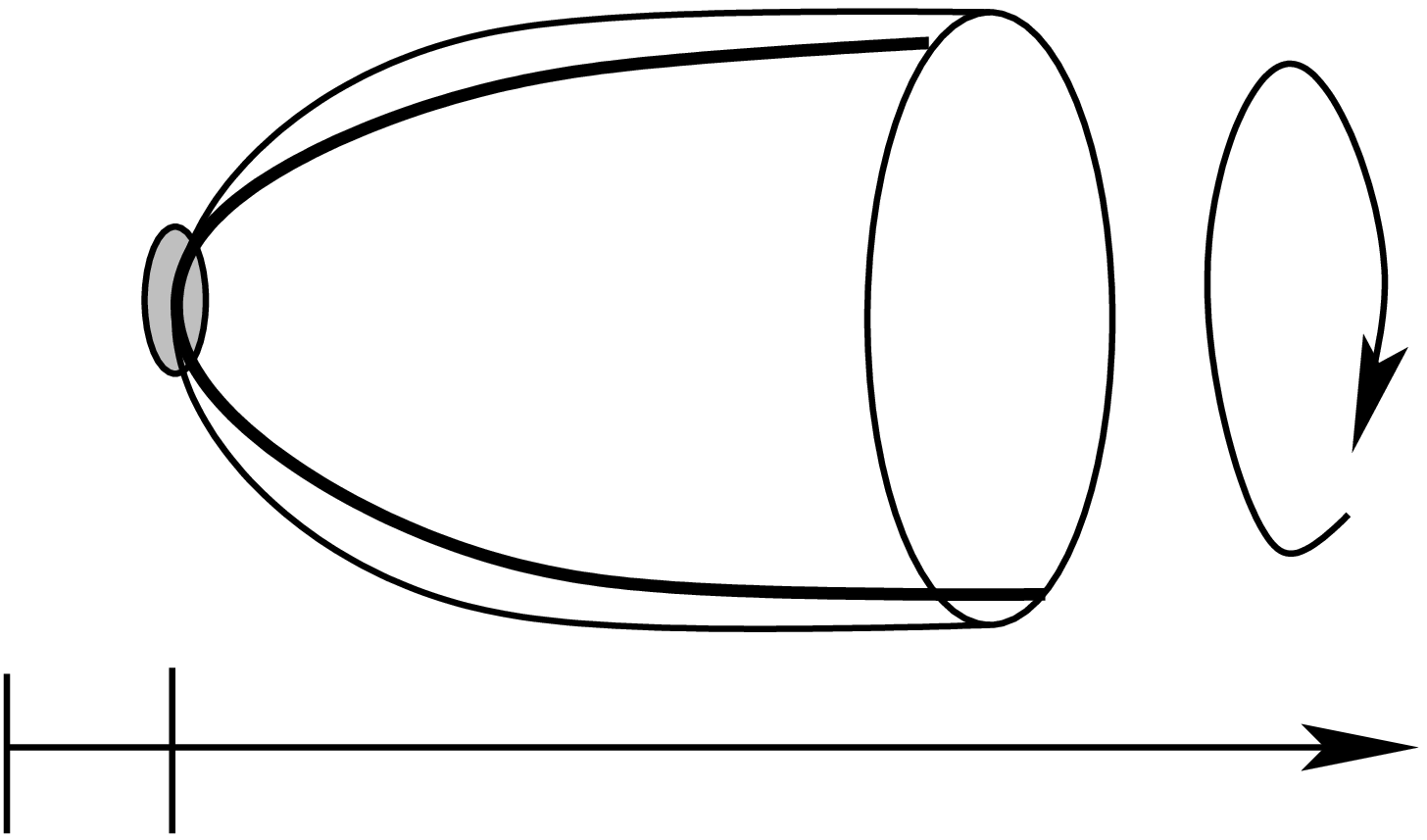}
\put(-200,-10){$0$}
\put(-180,-10){$U_{\rm KK}$}
\put(-20,-10){$U$}
\put(0,60){$\tau$}
\put(-120,95){D8s}
\put(-207,55){$A$ D4s}
\caption{Baryons are D4-branes sitting at the tip of the geometry
(depicted by a shaded blob).
They wrap $S^4$ which is not shown in this figure.}
\label{fig}
}
Collecting the baryon D4-branes, we consider $A$ of them to describe a
nucleus. On the $A$ D4-branes wrapping the $S^4$, one has 1+4
dimensional $U(A)$ gauge theory on ${\bf R}\times S^4$. At low energy of
these D4-branes, we have $U(A)$ gauge fields, 
adjoint scalar fields $\Phi_i$ ($i=1,2,3,y,z$)
and fundamental scalar fields $\rho_a$ ($a=1,2,\cdots, N_f$).
The adjoint and fundamental 
scalar fields are responsible for the location and the size of the
instantons, respectively.
The theory is a deformed version of ``ADHM'' gauge theory: it is
well-known that the theory on BPS D$p$-branes in the presence of  
D$(p+4)$-branes 
gives rise to ADHM equations of Yang-Mills instantons as BPS
equations\cite{Witten:1995gx,brane-within-brane}. 
In our case, the theory is not supersymmetric, due to 
the supersymmetry breaking in the bulk curved geometry. Therefore the
theory on the D4-branes has nontrivial potential terms which are
difficult to analyze.
We consider generic mass number $A$ for a nucleus. In particular when
$A$ is large, the theory is complicated.

The case of a single baryon was analyzed in \cite{HSSY}. In fact, there
appear two deformation terms; one is due to the background curved
geometry which gives rise to a potential for the $\Phi_z$ field, and the
other is due to the background 4-form flux providing the Chern-Simons
coupling. The size of the instanton $\rho_a$ obtains a potential from
both of these two effects. Larger $\Phi_z$ or larger $\rho_a$ cost
energy due to the background metric. So $\Phi_z$ is stabilized at $z=0$,
{\it i.e.} the instanton localizes at the tip of the geometry. 
When the size $\rho_a$ is small, electric charge (=instanton
number) density localizes and costs more energy, so the size is
stabilized at a certain finite value. This was computed in \cite{HSSY}
as
\begin{eqnarray}
 \rho = 3\sqrt{3\pi} \left(
\frac65\right)^{1/4} \frac{1}{\sqrt{\lambda}M_{\rm KK}}.
\label{rho}
\end{eqnarray}
Quantization of this electrically charged (=dyonic) 
instanton on the D8-branes was carried out in \cite{HSSY} 
to obtain a spectrum of baryons.

Normally, in holographic QCD, all the KK modes on $S^4$ are discarded,
simply because those excitation don't exist in QCD. If we apply the same
philosophy to the $U(A)$ gauge theory on the baryon D4-branes wrapping
the $S^4$, the gauge theory on the D4-branes reduce to a matrix quantum
mechanics. This theory is, in the holographic QCD, 
a microscopic theory of multi-body nucleons.

\section{Large $A$ limit and gravity dual}

Since the $U(A)$ gauge theory is difficult to treat, we take large $A$
limit and consider its dual geometry. 

Let us explain the whole picture first. Note that the $U(A)$ gauge
theory lives in the gravity dual of QCD, already. The $A$ D4-branes are
in the curved background created by the $N_c$ color D4-branes. 
In this situation, we consider a large $A$ limit. 
The large number of D4-branes
in the Witten's geometry back-reacts and deforms the geometry.
We take the near-horizon limit of this back-reacted geometry
by getting closer to the baryon D4-branes. So, effectively, we take
the AdS/CFT duality twice.\footnote{This is not rare situation.
For example, Coulomb branch of ${\cal N}=4$ super Yang-Mills was 
studied by a probe D3-brane in $AdS_5$, and if one collects many
D3-branes there, one needs to go to a back-reacted geometry
which is a multi-center D3-brane solution \cite{Kraus:1998hv}. 
What we are doing here is qualitatively similar to that.}

Before taking the near horizon limit, we need a supergravity solution.
In fact, it is not just a supergravity but the one with the flavor
D8-branes. The system consists of the supergravity in Witten's
background + $U(N_f)$ gauge theory on the D8-branes. We introduce $A$ 
dyonic instantons on the D8-brane gauge theory in the Witten's
background, 
and need to solve the equations of motion for 
the gravitating dyonic multi-instantons there.

This turns out to be very difficult, in various sense. First,
multi-instanton solutions with large instanton number are difficult to
construct (for example, see \cite{Brihaye:1982cc}). 
Second, the instanton equation has
two deformations described above. Thirdly, the instantons live only on
the D8-branes which are localized in the bulk, so the gravity problem is
essentially similar to that of braneworld black holes, which is known to
be difficult to solve even numerically. 
 
\FIGURE[t]{
\includegraphics[width=3cm]{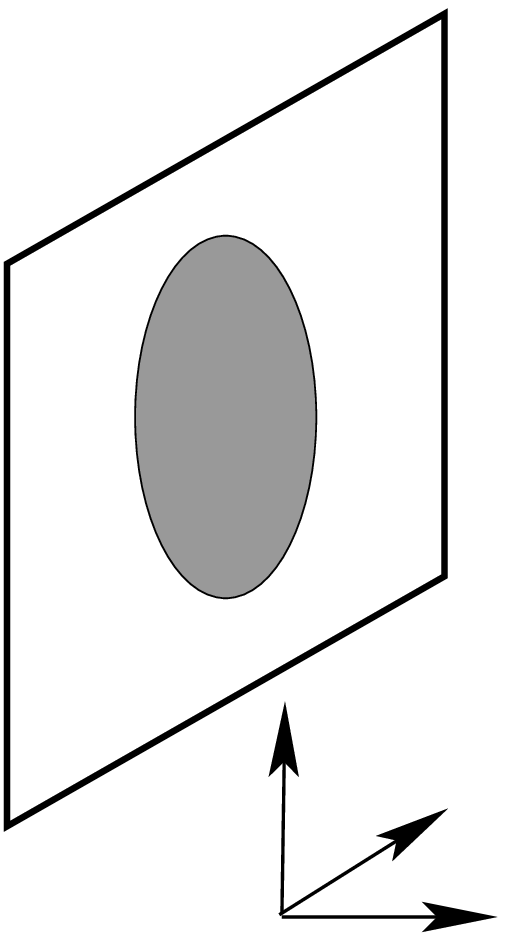}
\put(5,0){$y$}
\put(-30,30){$x^1,x^2,x^3$}
\caption{The 3-ball distribution of $A$ D4-branes (shaded region) on
D8-branes.}
\label{fig2}
}

So, we need to assume some gravitating dyonic instanton solution, to
proceed. For that, let us study expected property of the solution.
After all, we come to a conclusion that the solution is approximated by 
asymptotically-flat 
black-4-branes whose singularities are distributed on a 3-ball with 
a certain radius. See Fig.~\ref{fig2}.

First, let us recall a well-known property of nuclei. Among nucleons a
very strong repulsive force exists, while there is a binding force due
to light meson exchange. Due to this, basically nucleons 
are stabilized at a certain distance. Resultantly, any nucleus has a
common nucleon density, described by a nucleus radius formula
\begin{eqnarray}
 R_N(A) = \rho A^{1/3}.
\end{eqnarray}
Experimentally this is valid well, with 
$\rho \sim 1.1$[fm]. 
Interestingly, the value (\ref{rho}) obtained in
\cite{HSSY}, 
with $M_{\rm KK} \sim 500$ [MeV] which fits the baryon spectrum,
coincides with the experimental value of $\rho$. It should be possible
to reproduce this nuclear force from the SS model \cite{HSSp}, but here
we assume this homogeneous distribution of dyonic instantons over a
3-ball on the D8-brane. The radius of the 3-ball is, due to
(\ref{rho}),\footnote{If we use charge radius of proton (which was
obtained in the SS model in \cite{HSS}) instead of
(\ref{rho}), we have a different scaling in $\lambda$. The author thanks
S.~Sugimoto for pointing this out.}
\begin{eqnarray}
 R_N \sim \frac{A^{1/3} }{\sqrt{\lambda} M_{\rm KK}}.
\label{RN}
\end{eqnarray}
We neglect all numerical coefficients in this paper.
Note that the ball is in $(x^1,x^2,x^3)$ space, and the dyonic instantons 
are localized at $z=0$. These 4 directions are spatial directions of the
D8-branes except for the $S^4$ directions.
The dyonic instantons feel a potential along $z$ \cite{HSSY} 
and are stabilized at $z=0$. In fact
dyonic instantons with wave functions which expands along $z$ direction
correspond to excited nucleons
\cite{HSSY} which are not expected to be present in heavy nuclei.

Second, notice that the size of the nucleus $R_N$ (\ref{RN}) is small
for large $\lambda$. We are interested in the 
near horizon limit of the $A$ D4-brane geometry, so only a small region
of the bulk geometry is necessary.  
The D4-branes are at $z=0$, so, before the back-reaction, the geometry
is approximated by 
\begin{eqnarray}
ds^2 = \left(\frac{U_{\rm KK}}{R}\right)^{3/2}\!\!
\eta^{\mu\nu}dx^\mu dx^\nu 
 + 
\frac49\left(\frac{R}{U_{\rm KK}}\right)^{3/2} \!\!
(dy^2 + dz^2) + R^{3/2} U_{\rm KK}^{1/2}
d\Omega_4^2
.
\label{gd}
\end{eqnarray}
This is almost flat space, with rescaled variables 
$\tilde{x}^\mu= (U_{\rm KK}/R)^{3/4} x^\mu$ and 
$(\tilde{y},\tilde{z})= (2/3)(R/U_{\rm KK})^{3/4} (y,z)$.
We place $A$ D4-branes distributed in a 3-ball with its radius $R_N$
(\ref{RN}). If we neglect the presence of the D8-branes, the
curvature of the $S^4$ and also the background flux, the back-reacted
solution is just a multi-center BPS black D4-brane,
\begin{eqnarray}
&& ds^2 = f^{-1/2}
\left(-(d\tilde{x}^0)^2 + \lim_{R_{S^4}\to\infty} 
R_{S^4}^2 d\Omega_4^2\right) 
+ f^{1/2}
\left( (d\tilde{x}^i)^2 +  d\tilde{y}^2 +d\tilde{z}^2 \right),
\\
&& f \equiv 1 + \sum_{I=1}^A\frac{\pi g'_s l_s^3 }{r_I^3}, \quad
r_I \equiv |\vec{r}-\vec{w}_I|,
\label{f}
\end{eqnarray} 
with $\vec{r} = (\tilde{x}^i,\tilde{y},\tilde{z})$ ($i=1,2,3$),
and $\vec{w}_I$ are transverse location of the D4-branes (labeled by
$I=1,\cdots,A$).  
Here $g'_s$ is the string coupling constant evaluated at
$U=U_{\rm KK}$ (equivalently, $z=y=0$), 
\begin{eqnarray}
 g'_s = g_s \left(\frac{U_{\rm KK}}{R}\right)^{3/4}.
\end{eqnarray}
When the D4-brane distribution forms a uniform 3-ball of radius 
$\tilde{R}_N$, 
\begin{eqnarray}
 f = 1+ \frac{3A}{4\pi \tilde{R}_N^3} \int_{B^3} d^3\vec{w} \;
\frac{\pi g'_s l_s^3}{r_I^3}.
\end{eqnarray}

We need to take a near horizon limit. For this, it is necessary for $R_N$
to be smaller than ``would-be-AdS radius'' $R_0$ which is determined by
the factor in the harmonic function $f$ in (\ref{f}),
\begin{eqnarray}
 \tilde{R}_0 \equiv (\pi A g'_s)^{1/3} l_s, \quad R_0 = 
\tilde{R}_0/\sqrt{g_{ii}}.
\label{defr}
\end{eqnarray}
If the condition is violated, the ball is larger than the ``AdS
radius'' and thus AdS/CFT correspondence doesn't work.
In (\ref{defr}),
we have included the rescaling factor 
$\sqrt{g_{ii}} = (U_{\rm KK}/R)^{3/4}$ necessary for scaling 
$\tilde{x}$ back to $x$. Substituting (\ref{sugrap}), we obtain
\begin{eqnarray}
 R_0 = \frac32 \frac{A^{1/3}}{N_c^{1/3} M_{\rm KK}}.
\label{R0}
\end{eqnarray}
The condition $R_N< R_0$ is satisfied if 
\begin{eqnarray}
 N_c^{1/3} < \sqrt{\lambda}.
\label{req}
\end{eqnarray}
This can be satisfied for a chosen order of large $N_c$ and large
$\lambda$ limit.\footnote{It was pointed out in \cite{Kruczenski:2003uq}
(and argued also in \cite{SaSu1})
that the original supergravity approximation for the 
background (\ref{g}) is valid for 
$g_{\rm YM}^4 \ll \frac{1}{g_{\rm YM}^2 N_c}$. 
This contradicts 
(\ref{req}). However, note that these conditions are just for
order estimates which neglect numerical factors. 
In this paper we 
simply assume that there is a parameter region 
(for $g_{\rm YM} and N_c$) at which both the
above condition and the condition $R_N<R_0$ can be satisfied.
The author thanks H.~Ooguri and S.~Sugimoto
for bringing this discussion to him.} 
(If we substitute values used in \cite{SaSu1} which are
fixed by fitting the experimental $\rho$ meson mass and pion decay 
constant, this 
inequality is satisfied.) In this parameter region, the near horizon limit
makes sense.

\setcounter{footnote}{0}

\section{AdS/CFT dictionary and nuclear spectrum}

According to the 
standard AdS/CFT dictionary, gauge invariant operators of
the $U(A)$ gauge theory on the nucleus correspond to fluctuations of 
supergravity fields and fields on the flavor D8-branes, in the
background described in the previous section. More precisely, for
example, the dictionary consists of
\begin{eqnarray}
 {\rm tr}\left([\Phi_i,\Phi_k][\Phi_j,\Phi_k]\right) + \cdots
& \quad \leftrightarrow \quad & \delta g_{ij}
\nonumber \\
 \rho_a^\dagger \gamma_i \rho_b + \cdots
&\quad  \leftrightarrow \quad & \delta (A_i)_{\;\; b}^{a}
\nonumber
\end{eqnarray}
Fluctuation spectrum of the bulk fields
is the spectrum of these composite operators. In
particular, collective motion of the constituent nucleons is expressed
by these bulk fields. For example, 
the collective motion involving movement of the nucleons
corresponds to gravity fluctuations, while the gauge field fluctuation 
on the
D8-branes are mainly responsible for nucleon density waves in the
nucleus.  Bulk dilaton fluctuation includes all the effects.

We can perform a supergravity fluctuation analysis using the 
explicit metric given in the previous section. The simplest choice of
the fluctuation field is the dilaton fluctuation, which satisfies free
equation of motion in the curved geometry,\footnote{Generically
fluctuation of dilaton can mix with other gravity fluctuations. 
Here we are interested in only the order of gap of discrete
energy spectrum, as we will see below.}
\begin{eqnarray}
 \frac{1}{\sqrt{-g}} \partial_M e^{-2\phi}
\sqrt{-g} g^{MN}\partial_N \phi=0.
\end{eqnarray}
It turns out that, to solve this equation in the background black 4-brane
solution, one needs to do a numerical calculation, even if the presence
of the D8-branes is neglected.\footnote{Precise treatmnt is under
investigation \cite{KH}.}

What we want to see is if the spectrum is discrete or not, and if it is, 
what is the approximate gaps appearing in the spectrum. For this
purpose, we shall look at a similar black-brane system for which
extensive analyses have been performed already in the past. 
One very similar black brane solution is found in the supersymmetric
context, that is, ${\cal N}=4$ supersymmetric Yang-Mills theory
in Coulomb branch. In 
\cite{Freedman:1999gk,Brandhuber:1999jr,Gubser:1999eu} 
(see also \cite{Kraus:1998hv,Giddings:1999zu}), 
certain ball-like distributions of
D3-branes are considered. It was found there that dilaton fluctuation
develops a mass gap of order $R_N/R_0^2$ where in that case $R_N, R_0$
are ball radius and the AdS radius, respectively. In particular, it is
necessary for the ball dimension to be larger than 2, in order to have a
discrete spectrum \cite{Freedman:1999gk}. 
The typical scalar 
fluctuation spectrum \cite{Freedman:1999gk} is\footnote{In this paper we
consider Sakai-Sugimoto model, but we expect that starting from other
models may lead to similar results. For D3-D7 model 
\cite{Karch:2002sh} (or for cut-off $AdS_5$ with hard wall in bottom-up
approach \cite{Erlich:2005qh}), 
baryons are D5-branes wrapping $S^5$, so the heavy nucleus is dual to
near horizon geometry of $A$ D5-branes.} 
\begin{eqnarray}
 E_n \sim \frac{R_N}{R_0^2} \; n \quad \quad n=1,2,\cdots
\label{spe}
\end{eqnarray}
The striking point is that the continuous distribution of the D-branes
makes the spectrum discrete and gives a mass gap. The result is 
robust qualitatively, for 3- and 5-balls, and 3-sphere.\footnote{
For a 2-ball (a disk), there is a mass gap of order $R_N/R_0^2$ while
the spectrum is continuous above it \cite{Freedman:1999gk}.}

Interestingly, (\ref{spe}) has 
apparent similarity to the case of gravity
background for  
confining phase. For confining gauge theories, geometry is truncated at a
certain radius and one cannot go deeper beyond the wall, which gives
rise to the discrete spectrum. In the present
case, this wall is the ball surface.

Our background is made by the $A$ 
D4-branes while the above is for distributed 
D3-branes. However, there is an argument \cite{Gubser:1999eu} 
for qualitative understanding
of the spectrum formula (\ref{spe}). The denominator of (\ref{spe})
is $R_N$ which is naturally understood as energy coming from a string
stretched between the distributed D-branes. The average length is
proportional to $R_N$. In order to fix the dimensionality, one needs
$1/R_0^2$ in the formula.

Assuming this robustness of the formula (\ref{spe}) for distributed
D-branes in a ball, we can apply it to our particular case. If we
substitute (\ref{RN}) and (\ref{R0}), we obtain
\begin{eqnarray}
 E_n \sim \frac{N_c^{2/3}}{A^{1/3}\sqrt{\lambda}} M_{\rm KK} 
\; n
\quad \quad (n=1,2,\cdots)
\label{speg}
\end{eqnarray}
We neglected all the numerical factors. 
This (\ref{speg}) is the excitation spectrum of heavy nuclei with
mass number $A$, as a result of AdS/CFT correspondence.

It is surprising that this formula is actually consistent with 
the experimental
data of heavy nuclei, as follows. It is known 
from experiments that heavy nuclei have coherent excitations (phonons)
of constituent nucleons, called giant resonances,\footnote{
The author is indebted to T.~Nakatsukasa for invaluable discussions on
this correspondence, and also to T.~Matsui for his helpul explanations.}
\begin{eqnarray}
 E_n = \omega(A) \; n \quad\quad (n=1,2,\cdots)
\label{real}
\end{eqnarray}
Excitations with lower $n$ have been analysed in detail
experimentally. Phenomenological models, including liquid drop model of
heavy nuclei, indicates consistent harmonic behaviour in $n$.
 
Interestingly, the
harmonic-oscillator-like spectrum (\ref{real}) proportional to $n$
is similar to what we obtained,
(\ref{speg}), via gauge/gravity duality.
In the liquid drop model, the harmonic spectrum emerges from 
nuclear surface oscillation, so the spectrum is considered to arise from
collective motion of constituent nucleons. Our AdS/CFT interpretation,
due to the dictionary, is consistent with this.

Finally, let's compare 
numerical values of our formula with experiments. 
We are interested only in order of magnitudes.
The gap among discrete states in our formula is evaluated as 
\begin{eqnarray}
 \frac{R_N}{R_0^2} \sim 
\frac{N_c^{2/3}}{A^{1/3}\sqrt{\lambda}} M_{\rm KK}
\sim {\cal O}(10^2) \times A^{-1/3}\;{\rm [MeV]}
\label{res}
\end{eqnarray}
with typical values of the
parameters in the SS model, $M_{\rm KK} \sim 500$ [MeV],
$\lambda \sim {\cal O}(10) - {\cal O}(100)$ and $N_c=3$.
Experimental results for giant resonance of 
heavy nuclei with $A>60$ shows an empirical formula,
$E\sim 80 A^{-1/3}$ [MeV], 
for the first gap of $0^+$ isoscalar excitation.
We considered a dilaton fluctuation which corresponds to $0^+$,
and our result (\ref{res}) is not so much different from the nuclear
data. In particular, we reproduced the $A$ dependence of the resonance
excitations.  
As we adopted very crude
approximation (we discard all numerical factors and even adopted
similar but different geometry), one shouldn't take the numerical values 
seriously.

\section{Conclusion and discussion}

In this paper, we have given a dual gravity description of heavy atomic 
nuclei, by applying the gauge/string duality (AdS/CFT
correspondence). Since nucleons are described by D-branes wrapping
a sphere in 
curved geometry of holographic QCD, on a nucleus with mass number $A$
there appears a $U(A)$ gauge theory.
We took a large mass number limit $A \rightarrow \infty$. 
Dual gravity description is valid in this limit, and 
we obtained a near horizon geometry corresponding to the heavy nucleus.
The corresponding supergravity solution has discrete fluctuation
spectra, and we compared them with nuclear experimental data. 

Some discussions are in order.
We made some arguments for approximating the dual geometry, and it is
important to validate those by finding an explicit geometry
and performing fluctuation analysis. More precise treatment may reveal 
other aspects of the fluctuations. For example, 
we can put a probe D4-brane as an
additional nucleon, and consider a probe dynamics in this
near horizon geometry of $A$ D4-branes.
This may lead to a shell model potential of nuclear theory. 

Since our result (\ref{speg}) has an $A$ dependence, one can compare
it with $A$ dependence in experiments.
In the standard holographic QCD, $N$ of ``large $N$'' Maldacena limit is
fixed to be $N_c=3$ for comparison with nature. 
On the other hand, we have another number $A$
counting the nucleon D-branes, which appears in nature as a variety of
heavy nuclei. This enlarges possibility of comparison of AdS/CFT 
with nature.
Another aspect which can be distinguished from other
applications of AdS/CFT to nature is that clearly we have a 
$U(A)$ gauge theory on the CFT side. 
We don't rely on any universality for conformal physics.

Since the supersymmetries are completely broken, one may wonder if event
horizon may develop in the dual geometry and the dual gravity would
become a black hole. We don't expect it, because, 
if this were the case, the fluctuation spectrum
would become continuous (as known in deconfinement phase of
holographic QCD). 
However, 
since black holes are described by fluid
dynamics holographically \cite{fluid}, 
one can speculate that the liquid drop
model of heavy nuclei may be related to dual geometries through the
holographic hydrodynamics. 
In fact, dissipation of excitations on a nucleus is a target of 
research for many decades.\footnote{See for example \cite{Tomonaga}.
The author thanks M.~Natsuume and T.~Matsui for 
comments on this.} Investigation along this direction is in progress.

\acknowledgments 

K.~H.~would like to thank N.~Iizuka, G.~Mandal, T.~Matsui,
S.~Nakamura, T.~Nakatsukasa, M.~Natsuume, H.~Ooguri, T.~Sakai, 
S.~Sugimoto, T.~S.~Tai and T.~Yoneya for valuable discussions. 
K.~H.~is grateful to Haruko Hashimoto for a support on this project, 
and thanks Institute for the Physics and Mathematics of 
the Universe (IPMU)
for providing fruitful environment at Focus Week ``Quantum black
holes''.
K.~H.~is partly supported by
the Japan Ministry of Education, Culture, Sports, Science and
Technology. 

\newcommand{\J}[4]{{\it #1} {\bf #2} (#3) #4}
\newcommand{\andJ}[3]{{\bf #1} (#2) #3}
\newcommand{\AP}{Ann.\ Phys.\ (N.Y.)}
\newcommand{\MPL}{Mod.\ Phys.\ Lett.}
\newcommand{\NP}{Nucl.\ Phys.}
\newcommand{\PL}{Phys.\ Lett.}
\newcommand{\PR}{ Phys.\ Rev.}
\newcommand{\PRL}{Phys.\ Rev.\ Lett.}
\newcommand{\PTP}{Prog.\ Theor.\ Phys.}
\newcommand{\hep}[1]{{\tt hep-th/{#1}}}

\end{document}